%% file: acl_latex.tex
\documentclass[11pt]{article}

\usepackage[final]{acl}

\usepackage{times}
\usepackage{latexsym}

\usepackage[T1]{fontenc}

\usepackage[utf8]{inputenc}

\usepackage{microtype}

\usepackage{inconsolata}

\usepackage[most]{tcolorbox}
\tcbuselibrary{listings}
\usepackage{listings}
\usepackage{graphicx}
\usepackage{amsmath}
\usepackage{amssymb}
\usepackage{algorithm}
\usepackage{algpseudocode}
\usepackage{booktabs}
\usepackage{multirow}
\newcommand{\icon}{\raisebox{-0.2em}{\includegraphics[height=1em]{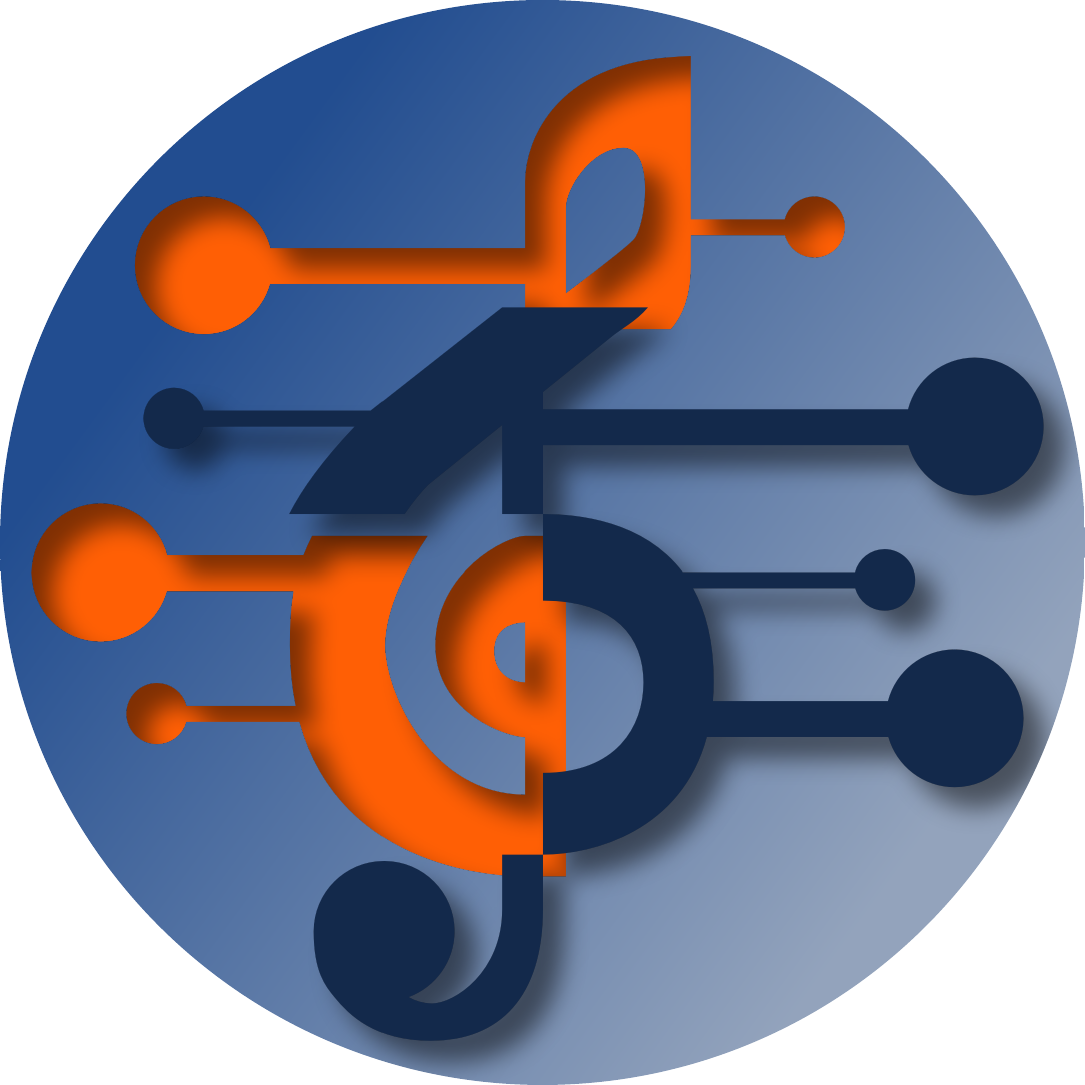}}}

\title{GBC: Gradient-Based Connections for Optimizing Multi-Agent Systems}

\author{
 \textbf{Xiaocheng Yang},
 \textbf{Abdulrahman Alrabah},
 \textbf{Dilek Hakkani-T\"{u}r},
 \textbf{Gokhan Tur}
\\[10pt]
 University of Illinois Urbana-Champaign
\\[5pt]
 \texttt{
  \{\href{mailto:xy61@illinois.edu}{xy61},
  \href{mailto:alrabah2@illinois.edu}{alrabah2},
  \href{mailto:dilek@illinois.edu}{dilek},
  \href{mailto:gokhan@illinois.edu}{gokhan}\}@illinois.edu
 }
}

\begin{document}
\maketitle
\begin{abstract}
Multi-agent systems (MAS) built on large language models (LLMs) provide a promising framework for solving complex tasks through role specialization and structured interaction. However, their performance is often limited by miscoordination and, more fundamentally, the lack of fine-grained credit assignment across agents. Existing approaches typically rely on coarse-grained feedback, making it difficult to identify which agents or interaction steps are responsible for errors. We propose Gradient-Based Connections (GBC), an approach for fine-grained attribution and optimization of multi-agent systems. GBC models a MAS as a computational graph and introduces gradient-based connection weights to quantify the influence of each agent’s output on downstream agents at the token level. By constructing an attribution graph and propagating task-specific loss signals backward, our method enables precise identification of error sources and targeted prompt optimization. We further develop AgentChord, an efficient implementation that leverages prefix-based gradient computation. Experiments on MultiWOZ and $\tau$-bench show that GBC improves multi-agent performance and outperforms strong single-agent and multi-agent baselines, and higher attribution quality is
associated with greater optimization effectiveness. Code is available at: \url{https://github.com/yxc-cyber/AgentChord}.
\end{abstract}

\input{sections/introduction}

\input{sections/related_work}

\input{sections/method}

\input{sections/framework}

\input{sections/experiment}

\input{sections/conclusion}

\input{sections/limitations}

\bibliography{custom}

\appendix

\input{sections/appendix}

\end{document}

%% file: sections/introduction.tex
\section{Introduction}

Large Language Models (LLMs) have enabled a new paradigm of multi-agent systems (MAS), where multiple specialized agents collaborate through structured interactions to solve complex tasks. By decomposing problems into sub-tasks and assigning them to different agents, multi-agent systems have demonstrated promise across a wide range of domains, including task-oriented dialogue, software engineering, and open-ended simulations \citep{gupta2024dardmultiagentapproachtaskoriented, wu2023autogenenablingnextgenllm, qian-etal-2024-chatdev}. However, despite their conceptual appeal, recent studies show that multi-agent systems often fail to consistently outperform strong single-agent baselines and suffer from issues such as miscoordination, inefficient communication, and lack of robust verification \citep{pan2025why}.

A fundamental challenge underlying these limitations is the lack of fine-grained credit assignment. In multi-agent workflows, errors in the final output often originate from specific agents or interaction steps, yet existing methods typically rely on coarse-grained signals (e.g., overall task success or reward) to guide optimization \citep{khattab2024dspy, xu2025metatextgrad, yuksekgonul2024textgradautomaticdifferentiationtext, 10.5555/3692070.3694667, luo2025agentlightningtrainai}. This makes it difficult to identify which components of the system are responsible for failures, limiting the effectiveness of both manual debugging and automatic optimization.

Recent advances in prompt optimization and LLM-based system design have explored gradient-inspired methods for improving performance, such as textual feedback propagation and self-supervised optimization \citep{yang2024largelanguagemodelsoptimizers, zhou2023large, pryzant-etal-2023-automatic, xiang-etal-2025-self-supervised}. While these approaches introduce more structured optimization signals, they are primarily designed for single-agent or monolithic pipelines, and do not explicitly address the challenges of attribution and optimization in multi-agent settings. In parallel, prior work on multi-agent systems has focused on improving coordination through architectural design—such as role specialization, graph-based communication, and task decomposition—but largely lacks principled mechanisms for token-level or interaction-level attribution across agents.

To address these challenges, we propose Gradient-Based Connections (GBC), a novel approach for optimizing multi-agent systems through fine-grained attribution. We model a multi-agent system as a directed computational graph and introduce a gradient-based connection mechanism that quantifies the influence of each agent’s output on subsequent agents at the token level. By constructing an attribution graph over agent interactions and propagating task-specific verbal loss signals backward through this graph, GBC enables precise identification of the components most responsible for errors. This facilitates more effective and targeted optimization of agent prompts.

Building on this formulation, we develop AgentChord, a practical framework that integrates GBC with a language-model-based optimizer to iteratively refine multi-agent systems. To ensure scalability, we introduce an efficient implementation that leverages a prefix-based gradient computation strategy to reduce memory overhead during backpropagation.

We evaluate our approach on both task-oriented dialogue (MultiWOZ~\citep{ye-etal-2022-multiwoz}) and interactive tool-use environments ($\tau$-bench~\citep{yao2025taubench}), demonstrating that GBC significantly improves multi-agent performance across multiple metrics. Our results show that fine-grained attribution enables effective optimization, and in some cases allows multi-agent systems to surpass strong single-agent baselines. Analysis of the MultiWOZ results further reveals an association between attribution quality and optimization effectiveness.

Our contributions are summarized as follows:
\begin{itemize}
    \item We propose Gradient-Based Connections (GBC), a method for token-level attribution across agents via gradient-based signals.
    \item We introduce AgentChord, a scalable framework for optimizing multi-agent systems using attribution-guided updates.
    \item We demonstrate the effectiveness of our approach on multiple benchmarks, including MultiWOZ and $\tau$-bench, showing improvements over single-agent and multi-agent baselines.
\end{itemize}

%% file: sections/related_work.tex
\section{Related Work}

\subsection{Prompt Optimization}

The performance of large language models (LLMs) is highly sensitive to prompt design, motivating extensive work on automatic prompt optimization. Early approaches formulate prompt design as black-box search over natural language instructions, where candidate prompts are generated and evaluated iteratively using LLMs themselves \citep{zhou2023large, yang2024largelanguagemodelsoptimizers}. These methods leverage LLMs as both generators and evaluators but rely on exploration guided by coarse-grained performance signals.

A complementary line of work introduces gradient-inspired optimization for prompts. ProTeGi models prompt refinement as following "textual gradients", where natural language feedback describing model errors is used to iteratively update prompts \citep{pryzant-etal-2023-automatic}. More generally, TextGrad extends this idea by treating LLM-based systems as computation graphs and propagating feedback signals across components, enabling automatic differentiation over prompt variables \citep{yuksekgonul2024textgradautomaticdifferentiationtext}. Recent work further explores self-supervised optimization. Self-Supervised Prompt Optimization (SPO) eliminates the need for labeled data by deriving optimization signals from pairwise comparisons of model outputs \citep{xiang-etal-2025-self-supervised}.

In addition, evolutionary approaches such as GAAPO and GEPA apply genetic algorithms to explore diverse prompt candidates through mutation and selection \citep{sécheresse2025gaapogeneticalgorithmicapplied, agrawal2026gepa}.

Despite these advances, these methods primarily focus on optimizing prompts for single-agent or single-step generation settings and rely on global performance feedback. They lack mechanisms for fine-grained attribution of errors to specific tokens, intermediate reasoning steps, or interacting components, limiting their effectiveness in complex multi-agent systems.

\subsection{Multi-Agent System}

Multi-agent systems (MAS) built on large language models (LLMs) have emerged as a powerful paradigm for solving complex tasks via role specialization, task decomposition, and iterative inter-agent communication. General-purpose frameworks such as AutoGen and ChatDev demonstrate that flexible multi-agent conversations and structured role-based collaboration can effectively coordinate LLM agents for complex workflows \citep{wu2023autogenenablingnextgenllm, qian-etal-2024-chatdev}. MAS have since been applied across diverse domains, including research simulation, open-ended social environments, and task-oriented dialogues \citep{yu2025research, ye2024an, gupta2024dardmultiagentapproachtaskoriented}.

To improve scalability and coordination, recent work introduces structured architectures for organizing agent interactions. Graph-based formulations represent agents and their communications as computational graphs, enabling systematic reasoning and optimization of information flow \citep{10.5555/3692070.3694667}. Similarly, DAG-based collaboration networks and task dependency graphs structure agent interactions for scalable coordination and complex task execution \citep{qian2025scaling, dong-etal-2024-villageragent}. Complementary approaches improve efficiency via role-aware routing and dynamic context selection \citep{liu2025rcrrouterefficientroleawarecontext}.

Despite these advances, recent studies show that MAS often provide limited gains over strong single-agent baselines and suffer from inter-agent misalignment, inefficient coordination, and weak verification \citep{pan2025why}. Moreover, identifying the source of errors remains challenging, as failures arise from specific agents or interaction steps but are difficult to attribute automatically \citep{zhang2025which}. These limitations are further compounded by broader risks such as miscoordination and emergent behaviors in complex multi-agent settings \citep{CAIF_1}.

\subsection{Multi-Agent System Optimization}

Beyond static architectures, a growing line of work treats multi-agent systems as optimizable programs. DSPy compiles declarative LLM pipelines into optimized execution graphs \citep{khattab2024dspy}, while TextGrad and metaTextGrad introduce gradient-like optimization mechanisms that propagate natural language feedback through computation graphs \citep{yuksekgonul2024textgradautomaticdifferentiationtext, xu2025metatextgrad}. In the multi-agent setting, GPTSwarm optimizes both agent behaviors and inter-agent connectivity within graph-structured systems, while MetaAgent automates the construction of agent organizations \citep{10.5555/3692070.3694667, zhang2025metaagent}. Reinforcement learning approaches further enable system-level optimization by learning from interaction trajectories and addressing credit assignment across agents \citep{luo2025agentlightningtrainai}.

Overall, existing approaches either rely on manually designed coordination mechanisms or optimize systems using coarse-grained signals, without providing fine-grained attribution across agents and interaction steps. In contrast, our work introduces a gradient-based connection framework that enables token-level attribution across agents, allowing more precise credit assignment and principled optimization of multi-agent systems.

%% file: sections/method.tex
\section{Method}
\label{sec:method}

\begin{figure*}[htp]
    \centering
    \includegraphics[width=1\linewidth]{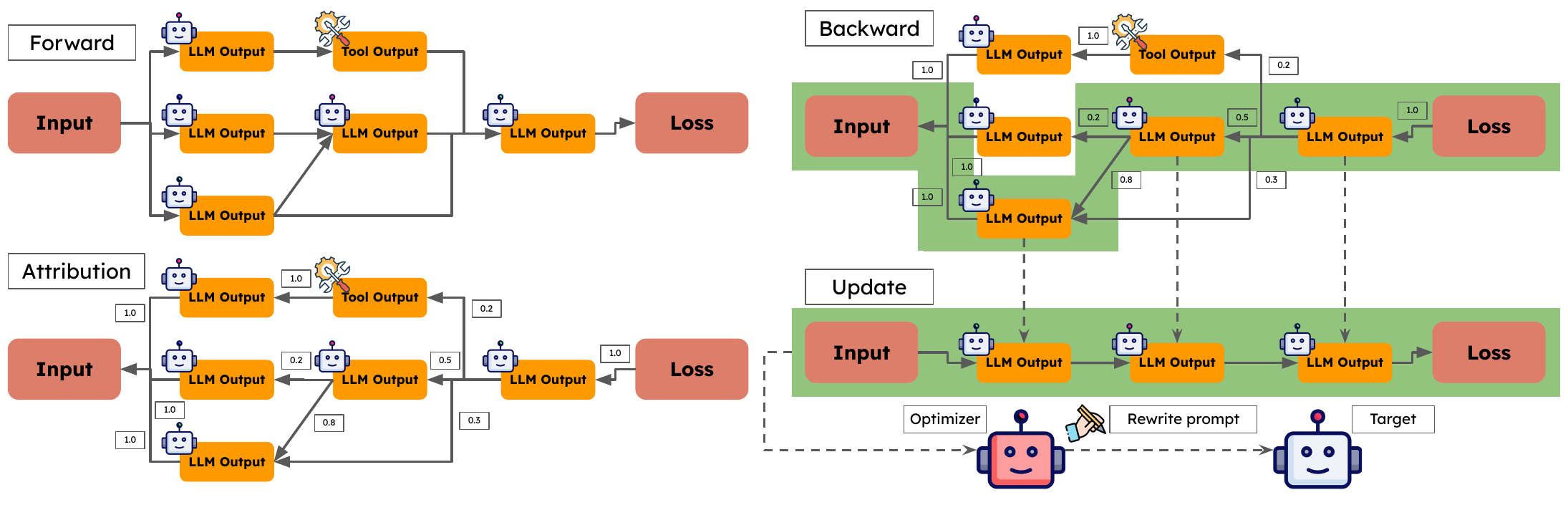}
    \caption{Overview of multi-agent system optimization with GBC. The procedure consists of four steps: \textbf{(1) Forward}, \textbf{(2) Attribution}, \textbf{(3) Backward}, and \textbf{(4) Update}. \textbf{(1) Forward}: The agent graph processes the input sequentially and produces the final output. \textbf{(2) Attribution}: Gradient-based connections construct an attribution graph that quantifies the influence of each predecessor output. \textbf{(3) Backward}: The framework propagates the loss backward through the attribution graph to extract attribution trajectories that identify the outputs most responsible for the final result. \textbf{(4) Update}: The optimizer updates agent prompts based on these trajectories to improve overall performance.}
    \label{fig:gbc horizontal}
\end{figure*}

We propose a framework for optimizing multi-agent systems with four components: agent graph, gradient-based connection, loss, and optimizer. Figure~\ref{fig:gbc horizontal} illustrates the overall pipeline. Given an input, the agent graph produces a final output through sequential agent interactions. Gradient-based connections construct an attribution graph that quantifies the influence of each predecessor output. A task-specific verbal loss is attached to the final output, and gradients are propagated backward to extract attribution trajectories identifying error sources. The optimizer then updates agent prompts based on these trajectories.

\subsection{Agent Graph}

We model the forward procedure of a multi-agent system as a directed acyclic graph $G = (V, E)$, following \citet{10.5555/3692070.3694667}. Each agent $v \in V$ is defined by a prompt–model pair $(P_v, M_v)$. Edges $E = \{(v_1, v_2) \mid O_{v_1} \subseteq I_{v_2}\}$ represent information flow.

The output of agent $v$ is:
\begin{equation}
O_v = M_v(P_v + I_v).
\label{equ:agent output}
\end{equation}

The input is:
\begin{equation}
I_v =
\begin{cases}
I_{\mathrm{initial}}, & \text{if } v \text{ is the first node}, \\
\sum_{u \in \mathrm{pre}(v)} O_u, & \text{otherwise}.
\end{cases}
\end{equation}

Agents are executed in topological order, and the final output is produced by the last agent.

\subsection{Gradient-Based Connection}

We introduce gradient-based connections to quantify the contribution of each predecessor output. For each $O_v$, we compute connection weights $W_{O_v}(O_u)$ for $u \in \mathrm{pre}(v)$ using gradient-based signals. We consider four variants: mean/max of L1 norm and mean/max of gradient–input product (Sections~\ref{sec:mean of l1 norm}–\ref{sec:max of product with input}).

For each output, we retain the top-$m$ predecessors to construct an attribution graph $G_{\text{attr}} = (V_{\text{attr}}, E_{\text{attr}})$:
\begin{equation}
\begin{aligned}
E_{\text{attr}} =
\{ &(v_{\text{attr},1}, v_{\text{attr},2}) \mid (v_{\text{attr},1}, v_{\text{attr},2}) \in V_{\text{attr}}^2, \\
& W_{O_{v_2}}(O_{v_1}) \in \mathrm{Top}_m(W_{O_{v_2}}) \}.
\end{aligned}
\end{equation}
We use $m=1$ by default.

\subsubsection{Mean of L1 Norm}
\label{sec:mean of l1 norm}

{
\small
\begin{equation}
W_{O_v}(O_u) =
\operatorname{avg}\!\Big(
\big\|
\nabla \prod_{w \in O_v}
\mathbb{P}(w \mid \mathrm{Embed}(O_u))
\big\|_{L_1}
\Big).
\end{equation}
}

This computes token-level salience scores via gradients and aggregates them by averaging.

\subsubsection{Max of L1 Norm}
\label{sec:max of l1 norm}

{
\small
\begin{equation}
W_{O_v}(O_u) =
\operatorname{max}\!\Big(
\big\|
\nabla \prod_{w \in O_v}
\mathbb{P}(w \mid \mathrm{Embed}(O_u))
\big\|_{L_1}
\Big).
\end{equation}
}

This emphasizes the most influential tokens, reducing noise from irrelevant ones.

\subsubsection{Mean of Product with Input}
\label{sec:mean of product with input}

{
\small
\begin{equation}
\begin{aligned}
W_{O_v}(O_u)
&= \operatorname{avg}\!\Big(
\nabla \prod_{w \in O_v}
\mathbb{P}(w \mid \mathrm{Embed}(O_u)) \\
&\quad \cdot \mathrm{Embed}(O_u)
\Big).
\end{aligned}
\end{equation}
}

This captures first-order contributions via gradient–input interactions \citep{10.5555/3305890.3306006}.

\subsubsection{Max of Product with Input}
\label{sec:max of product with input}

{
\small
\begin{equation}
\begin{aligned}
W_{O_v}(O_u)
&= \operatorname{max}\!\Big(
\nabla \prod_{w \in O_v}
\mathbb{P}(w \mid \mathrm{Embed}(O_u)) \\
&\quad \cdot \mathrm{Embed}(O_u)
\Big).
\end{aligned}
\end{equation}
}

This combines the gradient–input interactions with the emphasis on most influential tokens.

\subsection{Loss}

We define a task-specific verbal loss based on the system output and attach it to the attribution graph. The loss encodes correctness and quality signals, and can include fine-grained feedback (e.g., ground truth comparisons or explanations) to guide optimization. Details are provided in Appendix~\ref{app:loss}.

\subsection{Optimizer}

We backpropagate the loss through the attribution graph to obtain attribution trajectories:
\[
\tau = [(s_0, c_0), \dots, (\ell, L_\ell)].
\]

Each trajectory traces the contribution from inputs or intermediate outputs to the loss. The set $\mathcal{T}(\text{input})$ collects all such trajectories. The procedure is summarized in Algorithm~\ref{alg:backpropagation}.

\input{algorithms/backpropagation}

We then use a language model as the optimizer \citep{yang2024largelanguagemodelsoptimizers}. Given current prompts, attribution trajectories, and optimization history, it updates agent prompts to improve performance. Details are provided in Appendix~\ref{app:opro}.

%% file: algorithms/backpropagation.tex
\begin{algorithm}[h]
\small
\caption{Backpropagation of Attribution Trajectories}
\label{alg:backprop_attr}
\begin{algorithmic}[1]
\Require Attribution graph $G_{\mathrm{attr}} = (V_{\mathrm{attr}}, E_{\mathrm{attr}})$
\Require Loss node set $V_{\mathrm{loss}} = \{(\ell, L_\ell)\}$
\Require Initial input $I_{\mathrm{initial}}$
\State Initialize an empty list $\mathcal{T}(s)$ for each subject $s \in V \cup \{I_{\mathrm{initial}}\}$

\ForAll{$(\ell, L_\ell) \in V_{\mathrm{loss}}$}
    \State $\mathrm{cache} \gets [[(\ell, L_\ell)]]$
    \ForAll{$(v, O_v) \in \mathrm{pre}_{\mathrm{attr}}(\ell, L_\ell)$}
        \State \Call{Backward}{$(v, O_v), \mathrm{cache}$}
    \EndFor
\EndFor

\State \Return $\{\mathcal{T}(s)\}_{s \in V \cup \{I_{\mathrm{initial}}\}}$

\Function{Backward}{$(v, O_v), \mathrm{cache}$}
    \State $\mathrm{new\_cache} \gets [\ ]$
    \ForAll{$\tau \in \mathrm{cache}$}
        \State $\tau' \gets \mathrm{copy}(\tau)$
        \State Insert $(v, O_v)$ at the beginning of $\tau'$
        \State Append $\tau'$ to $\mathrm{new\_cache}$
    \EndFor
    \State $\mathcal{T}(v) \gets \mathcal{T}(v) \cup \mathrm{new\_cache}$

    \If{$\mathrm{pre}_{\mathrm{attr}}(v, O_v) = \emptyset$}
        \State $\mathrm{input\_cache} \gets [\ ]$
        \ForAll{$\tau \in \mathrm{new\_cache}$}
            \State $\tau' \gets \mathrm{copy}(\tau)$
            \State Insert $(input, I_{\mathrm{initial}})$ at the beginning of $\tau'$
            \State Append $\tau'$ to $\mathrm{input\_cache}$
        \EndFor
        \State $\mathcal{T}(input) \gets \mathcal{T}(input) \cup \mathrm{input\_cache}$
    \Else
        \ForAll{$(u, O_u) \in \mathrm{pre}_{\mathrm{attr}}(v, O_v)$}
            \State \Call{Backward}{$(u, O_u), \mathrm{new\_cache}$}
        \EndFor
    \EndIf
\EndFunction
\end{algorithmic}
\label{alg:backpropagation}
\end{algorithm}

%% file: sections/framework.tex
\section{AgentChord \texorpdfstring{\icon}{[icon]}}

Following the pipeline in Section~\ref{sec:method}, we develop \textbf{AgentChord}~\icon\footnote{Code is available at: \url{https://github.com/yxc-cyber/AgentChord}.}, a practical framework for multi-agent prompt optimization using Gradient-Based Connections (GBC).

To enable scalability, we introduce a \emph{prefix-based gradient computation} technique to reduce memory overhead. As defined in Equation~\ref{equ:agent output}, each agent processes a prompt and an input. Since attribution is computed only with respect to the input, gradients are required only for input tokens, while prompt tokens are treated as a fixed prefix.

In practice, we first pass the prompt through the model without gradients to obtain the KV cache, and then process the input with gradients enabled. This avoids storing gradients for prompt tokens.

As a result, memory complexity is reduced from
\[
\mathcal{O}(n \cdot d \cdot L)
\]
to
\[
\mathcal{O}((n - k) \cdot d \cdot L),
\]
where $n$ is the total sequence length, $k$ is the prompt length, $d$ is the hidden dimension, and $L$ is the number of layers.

%% file: sections/experiment.tex
\section{Experiment}

Multi-agent systems enable task decomposition and specialization, which can improve performance in complex, multi-domain settings. We evaluate GBC on two benchmarks: MultiWOZ for task-oriented dialogue and $\tau$-bench for interactive tool-use.

\subsection{MultiWOZ}

\begin{figure}[h]
    \centering
    \includegraphics[width=1\linewidth]{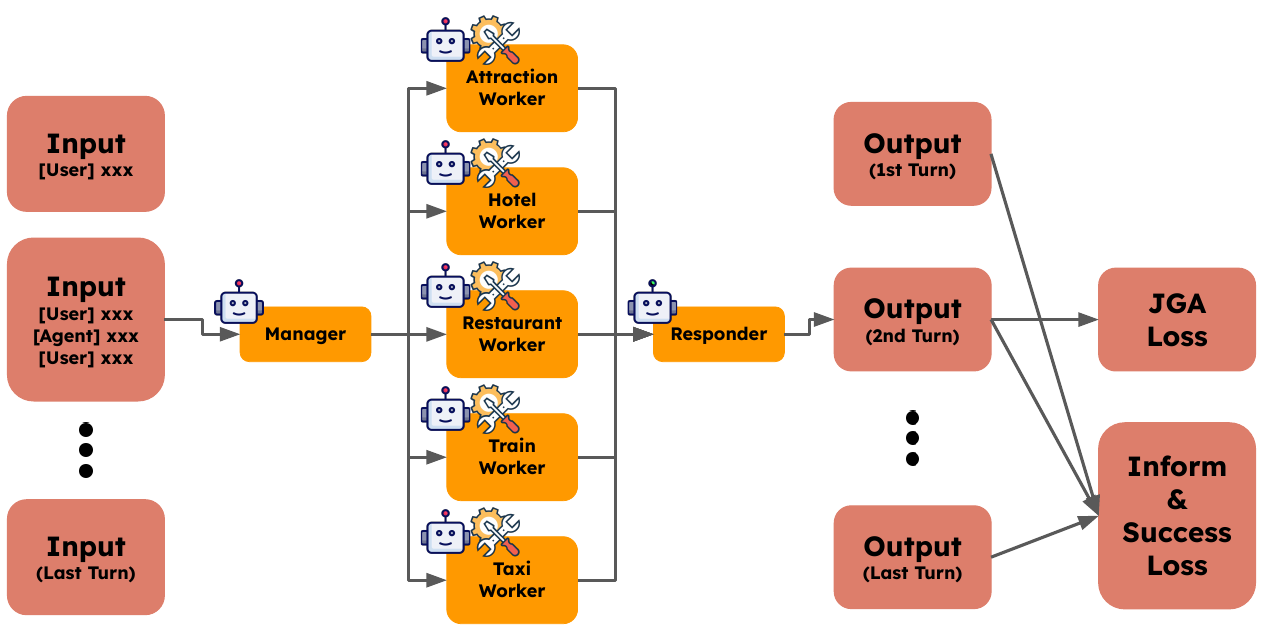}
    \caption{Multi-agent system tailored to MultiWOZ.}
    \label{fig:multiwoz framework}
\end{figure}

\input{tables/multiwoz_main}

\begin{figure*}[t]
    \centering
    \includegraphics[width=1\linewidth]{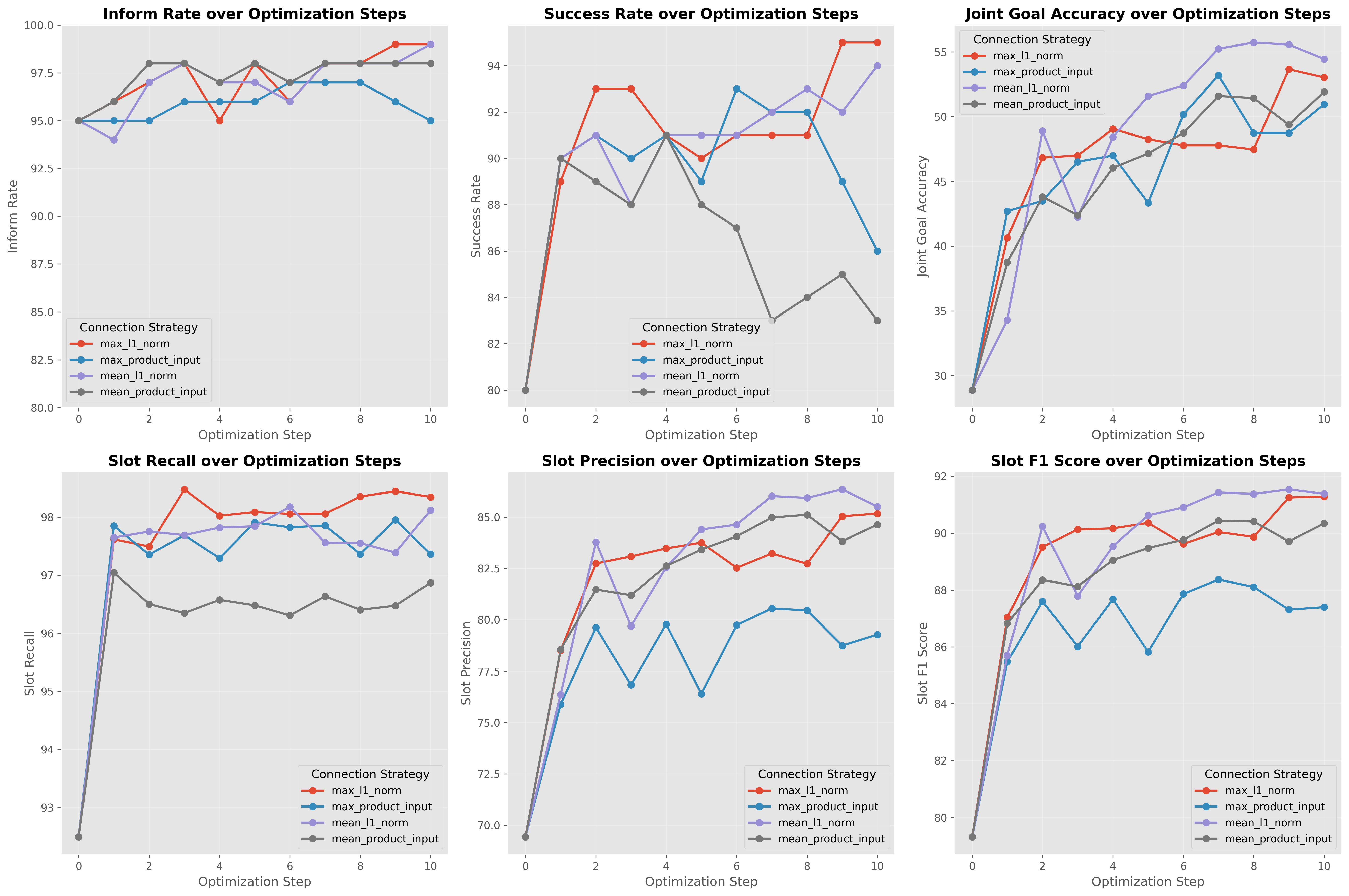}
    \caption{Optimization dynamics of Qwen-3-32B on MultiWOZ. Step 0 denotes the unoptimized multi-agent baseline. \textbf{Inform} and \textbf{Slot Recall} remain high throughout optimization, while \textbf{JGA}, \textbf{Slot Precision}, and \textbf{Slot F1} show clear upward trends, indicating improved dialogue-state tracking and fewer over-predicted slots. \textbf{Success} is more variable, suggesting that full goal completion remains harder to optimize. Overall, L1-norm-based connection weights achieve the strongest final performance.}
    \label{fig:optimization curve}
\end{figure*}

\paragraph{Setup}
MultiWOZ \citep{ye-etal-2022-multiwoz, budzianowski-etal-2018-multiwoz, eric-etal-2020-multiwoz, zang-etal-2020-multiwoz, 10.1007/978-3-030-88483-3_16} is a task-oriented dialogue benchmark with annotated dialogue states. We use MultiWOZ 2.4 and sample 100 dialogues from five domains (Attraction, Hotel, Restaurant, Train, Taxi). 

We adopt a manager–worker architecture (Figure~\ref{fig:multiwoz framework}). The manager assigns tasks based on dialogue context, domain-specific workers perform API calls, and a responder generates the final response.

\paragraph{Verbal Loss}
We use two types of verbal loss: (1) a turn-level JGA loss that compares predicted and ground-truth dialogue states, including false positive and false negative slot-value pairs; and (2) a dialogue-level Inform \& Success loss that evaluates whether the system retrieves correct entities and provides requested information. Detailed prompts are provided in Appendix~\ref{app:multiwoz loss}.

\paragraph{Metrics}
We report Inform, Success (conversation-level), and Joint Goal Accuracy (JGA), Slot Recall, Slot Precision, and Slot F1 (turn/slot-level).

\paragraph{Optimization Setup}
We use 30 training samples, updating prompts every 3 samples (10 steps total). Backbone models are Llama-3.3-70B-It and Qwen-3-32B.

\paragraph{Results}
Results are shown in Table 1. Before optimization, the multi-agent systems do not consistently outperform their single-agent counterparts. For example, although the Qwen-3-32B multi-agent system achieves higher Inform and Success scores than the single-agent baseline, its JGA and Slot F1 are substantially lower. After optimization with GBC, performance improves across most metrics, especially for Qwen-3-32B. With mean of L1 norm, Qwen-3-32B reaches the best overall MultiWOZ performance, improving JGA from 28.9 to 54.4 and Slot F1 from 79.3 to 91.4, while also achieving 99.0 Inform and 94.0 Success. The max of L1 norm variant obtains similarly strong results, with 99.0 Inform, 95.0 Success, 53.0 JGA, and 91.3 Slot F1. These optimized multi-agent systems substantially outperform the Qwen-3-32B single-agent baseline on Inform, Success, JGA, and Slot F1, demonstrating that GBC can convert an initially under-optimized multi-agent system into a stronger task-oriented dialogue agent.

Figure 3 further illustrates the optimization dynamics for Qwen-3-32B. Across the four connection weight formulations, most metrics, including Inform, JGA, Slot Recall, Slot Precision, and Slot F1, show clear upward trends over optimization steps. In particular, JGA and Slot F1 improve steadily from the early steps, indicating that attribution-guided prompt updates help the system better track dialogue states and predict slot values. Success remains more variable and challenging, suggesting that completing the full user goal still depends on difficult long-horizon coordination. Overall, the trends in Figure 3 are consistent with Table 1 and show that GBC provides stable improvements during optimization, with the L1-norm-based variants yielding the strongest final performance.

\begin{figure}[h]
    \centering
    \includegraphics[width=1\linewidth]{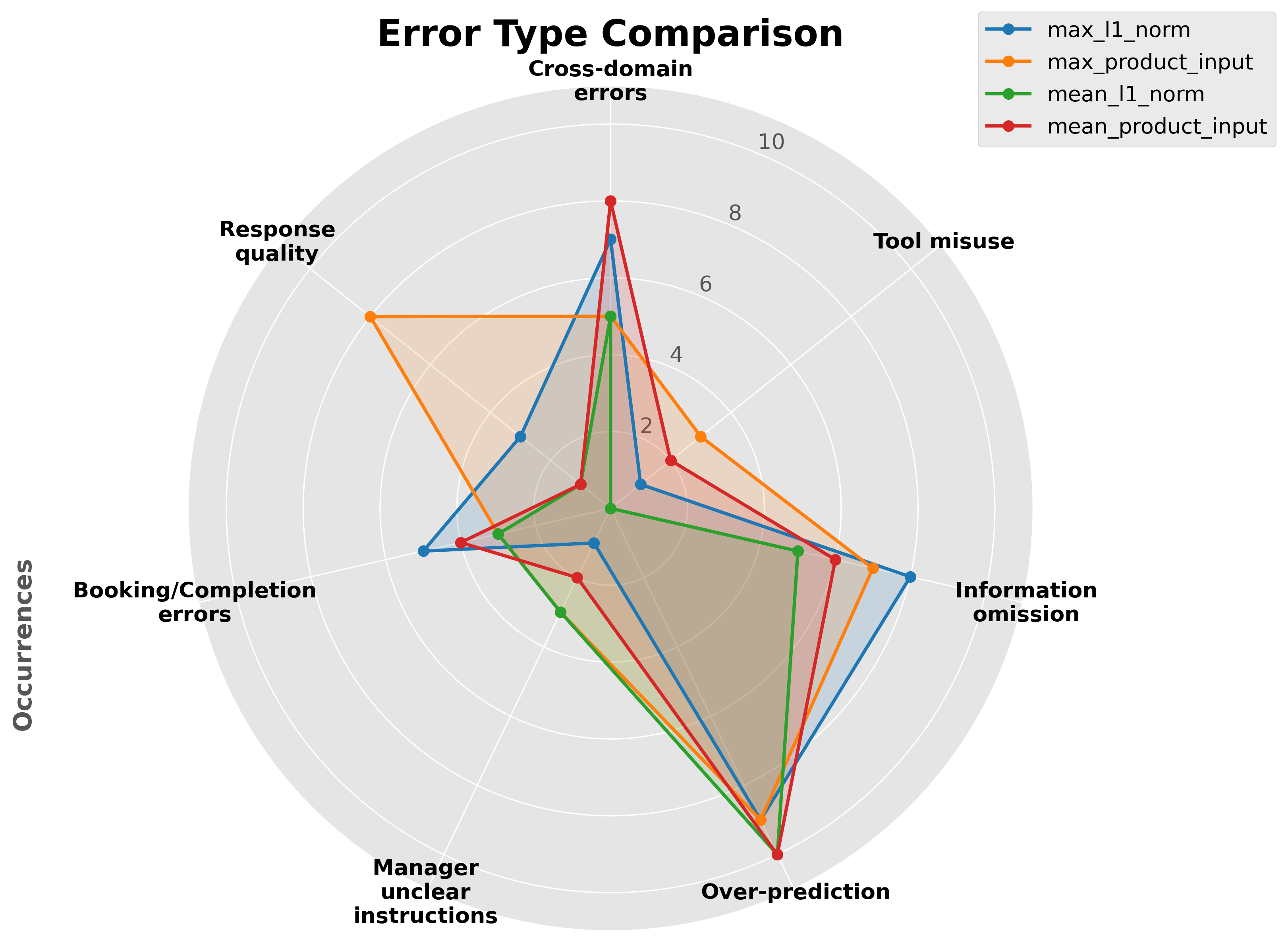}
    \caption{The occurrences of different error types detected by the optimizer under different connection weight formulae on MultiWOZ with Qwen-3-32B.}
    \label{fig:multiwoz error analysis}
\end{figure}

\paragraph{Error Analysis}
We categorize errors into seven types: cross-domain errors, tool misuse, information omission, over-prediction, unclear manager instructions, booking errors, and response quality issues. 

As shown in Figure~\ref{fig:multiwoz error analysis}, cross-domain errors, information omission, and over-prediction occur most frequently. This suggests that MultiWOZ failures are mainly caused by multi-domain coordination and dialogue-state tracking, rather than only by surface-level response generation. Cross-domain errors indicate that the manager-worker routing is still imperfect, especially when multiple domain-specific agents are available. Information omission shows that even when relevant information appears in the dialogue, it may be lost during extraction or inter-agent communication. Over-prediction further suggests that agents sometimes infer slot values beyond the evidence provided by the user. These patterns are consistent with the improvements in JGA, Slot Precision, and Slot F1 after optimization, since reducing omitted and over-predicted slots directly improves dialogue-state tracking. Overall, the error distribution indicates that GBC improves the system by targeting coordination and state-tracking failures, while cross-domain responsibility assignment remains a key challenge.

\begin{figure}[h]
    \centering
    \includegraphics[width=1\linewidth]{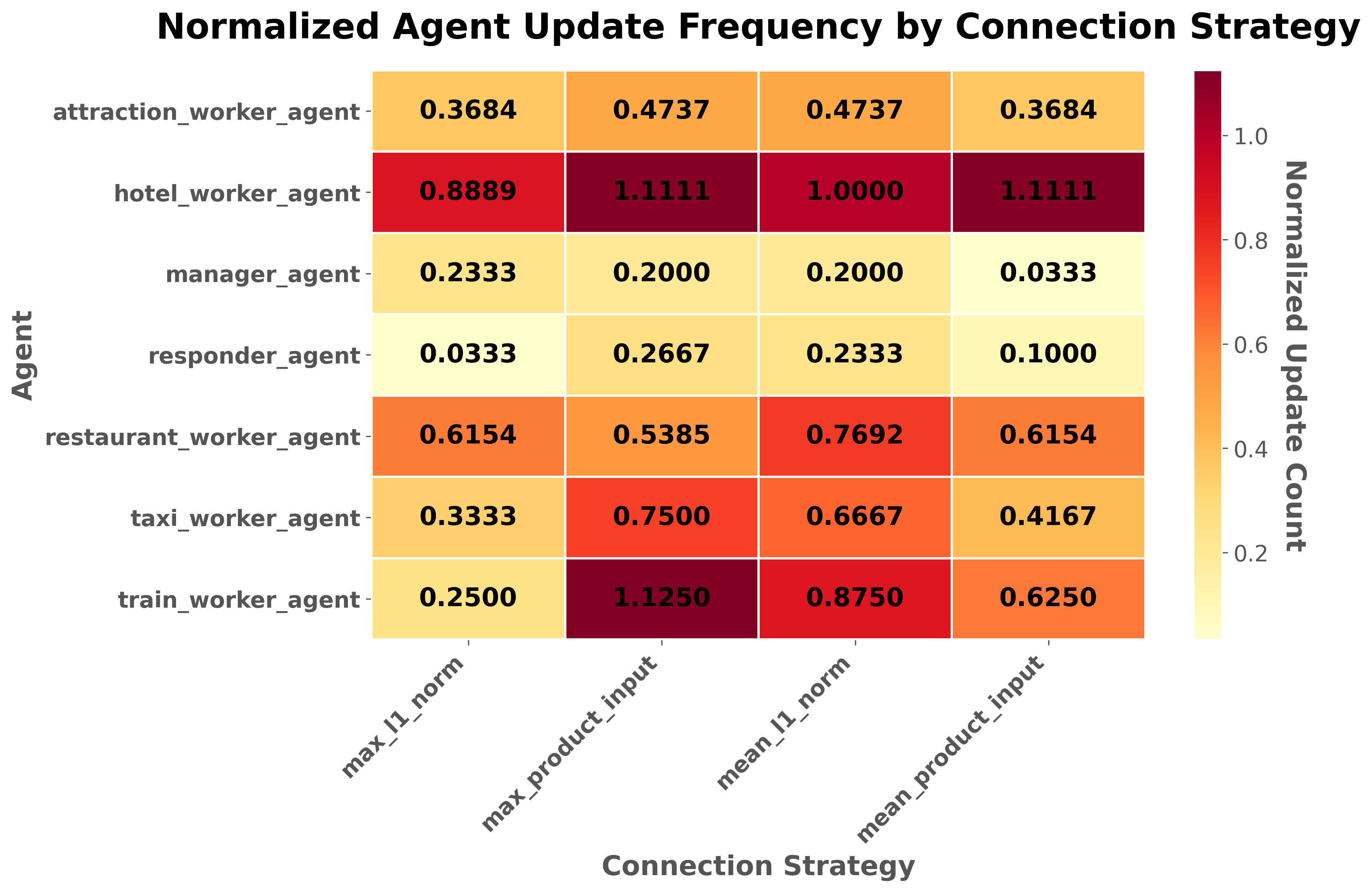}
    \caption{Heatmap of normalized agent update frequency by connection strategy with Qwen-3-32B.}
    \label{fig:multiwoz update}
\end{figure}

\paragraph{Update Analysis}
Figure~\ref{fig:multiwoz update} shows normalized update frequency (NUF):
\begin{equation}
\mathrm{NUF}(v) = \frac{|\{i \mid v \in U_i\}|}{|\{i \mid v \in R_i\}|}.
\end{equation}
$U_i$ is the set of updated agents for the $i$-th round and $R_i$ is the set of relevant agents for the $i$-th round. Responder and manager agents are always relevant, while domain-specific workers are relevant if and only if the task domain matches the work's domain.

Domain-specific workers are updated more frequently than manager or responder agents, consistent with the observed error patterns.

\begin{figure}[h]
    \centering
    \includegraphics[width=1\linewidth]{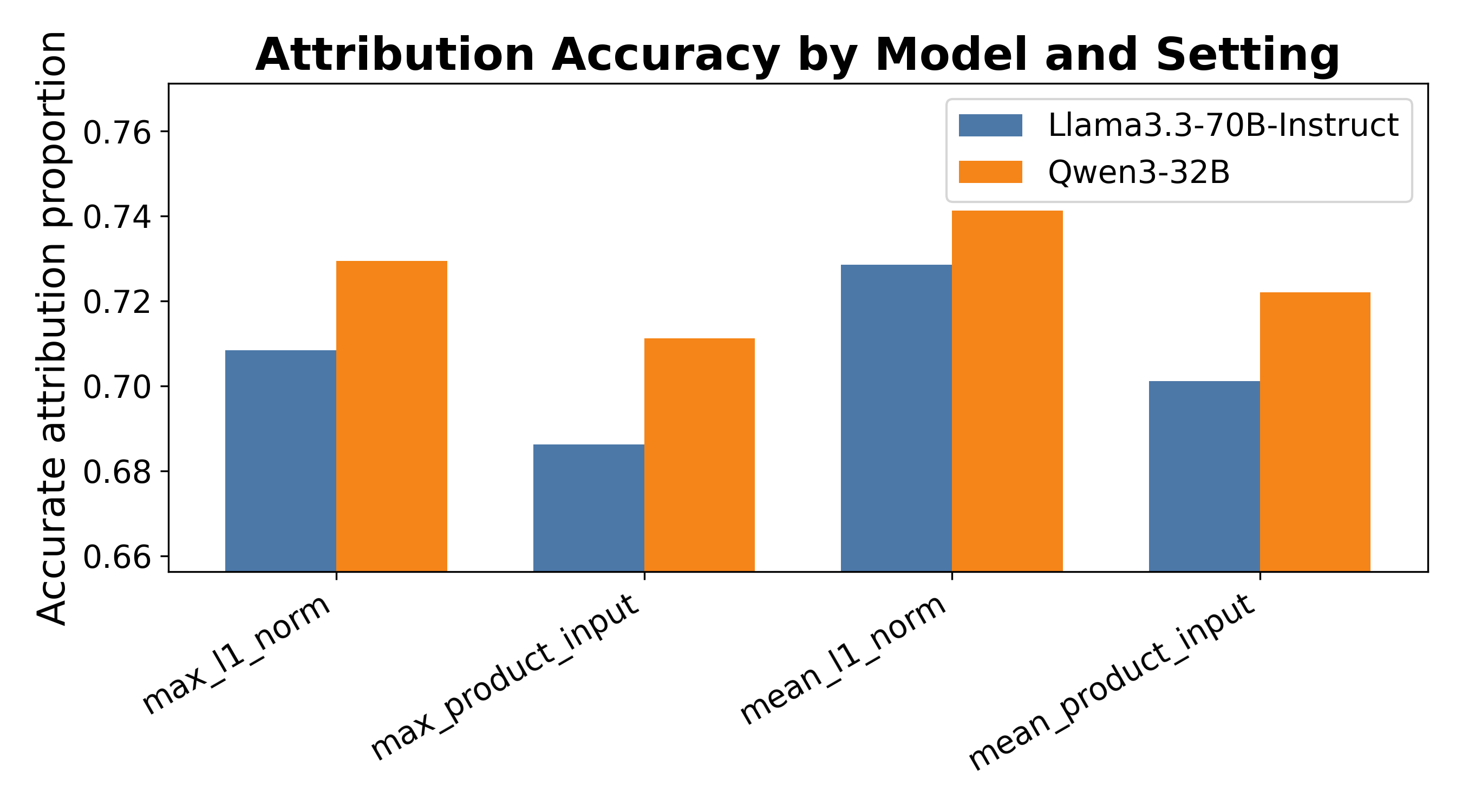}
    \caption{Attribution accuracy grouped by model within each connection weight setting.}
    \label{fig:multiwoz attribution accuracy}
\end{figure}

\paragraph{Attribution Quality Analysis}
We approximate the accuracy of attribution by checking whether each attribution trajectory contains the worker agents responsible for the domains of the dialogue. Figure~\ref{fig:multiwoz attribution accuracy} shows the attribution accuracy for the 2 models and 4 connection weights. Regardless of the model, mean and max L1 norm always lead to the best two attribution accuracies, which explains why those two connection weight formulae perform best in terms of the metrics of the MultiWOZ task. This reveals that higher attribution quality is associated with greater optimization effectiveness.

\subsection{$\tau$-bench}

\paragraph{Setup}
$\tau$-bench \citep{yao2025taubench} evaluates agents in multi-step tool-use environments. We focus on the retail domain due to the availability of training data.

\begin{figure}[h]
    \centering
    \includegraphics[width=1\linewidth]{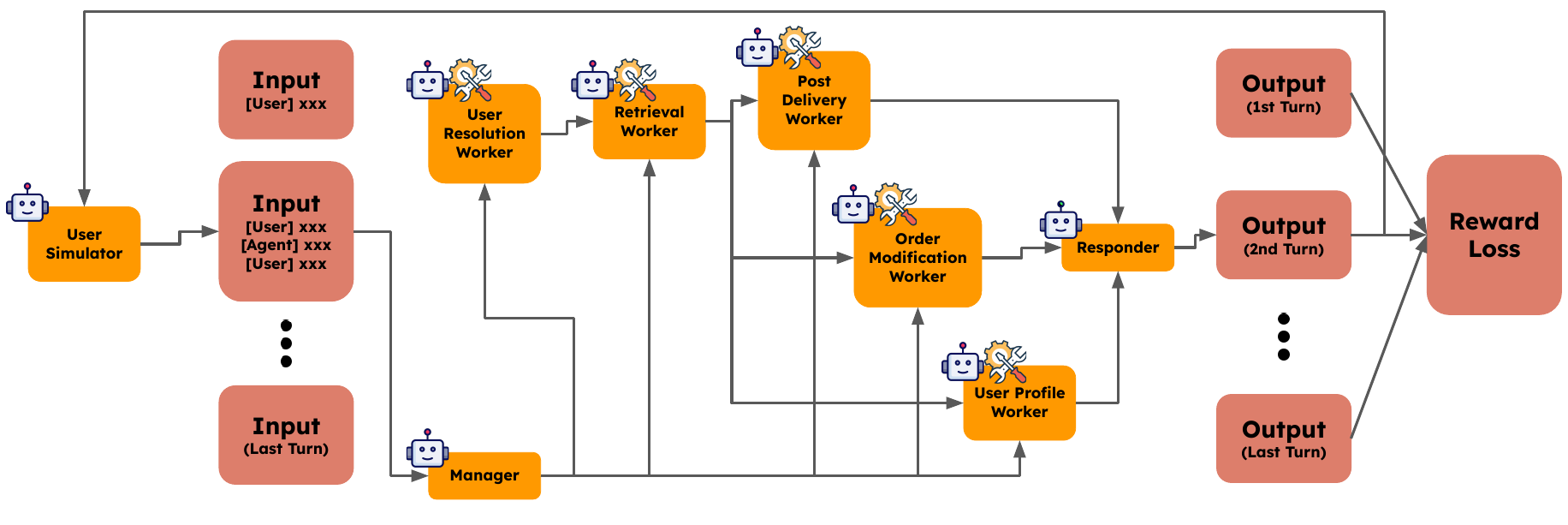}
    \caption{Multi-agent system tailored to $\tau$-bench.}
    \label{fig:tau-bench framework}
\end{figure}

The system follows a manager–worker design (Figure~\ref{fig:tau-bench framework}). Workers handle user resolution, retrieval, order modification, post-delivery, and user profile tasks, while a responder generates outputs. A user simulator interacts with the system iteratively.

\paragraph{Verbal Loss}
We define a reward-based verbal loss at the conversation level, which includes the ground-truth tool-call trajectory, the agent-generated trajectory, and required response contents. The loss evaluates both tool-call correctness and whether system outputs contain all required information. Details are provided in Appendix~\ref{app:reward loss}.

\paragraph{Metrics}
We evaluate performance using three conversation-level metrics. Action reward measures whether the sequence of tool calls matches the ground-truth trajectory. Output reward measures whether system responses contain all required information. Overall reward is defined as the product of the two, requiring both correct actions and complete responses.

\paragraph{Optimization Setup}
We use 10 training tasks and update prompts after each conversation. GPT-4o-mini is used as the user simulator due to budget constraints.

\input{tables/taubench_main}

\paragraph{Results}
Results are shown in Table~\ref{tab:taubench_main}. Optimization consistently improves multi-agent performance over its pre-optimization baseline. Among the optimized variants, max of L1 norm yields the strongest overall performance for Qwen-3-32B, improving overall reward from 13.0 to 24.3, surpassing the strong single-agent baseline of 22.6. Mean of product with input also reaches an overall reward of 24.3 for Qwen-3-32B, mainly driven by a large improvement in action reward from 13.9 to 27.0. For Llama-3.3-70B-It, all optimized multi-agent variants improve over the pre-optimization baseline, with max of product with input achieving the best overall reward, increasing from 6.1 to 9.6. These results demonstrate the overall effectiveness of GBC for optimizing multi-agent systems, as it consistently improves overall reward across both backbone models and enables the Qwen-3-32B multi-agent system to outperform its strong single-agent counterpart.

\begin{figure}[h]
    \centering
    \includegraphics[width=1\linewidth]{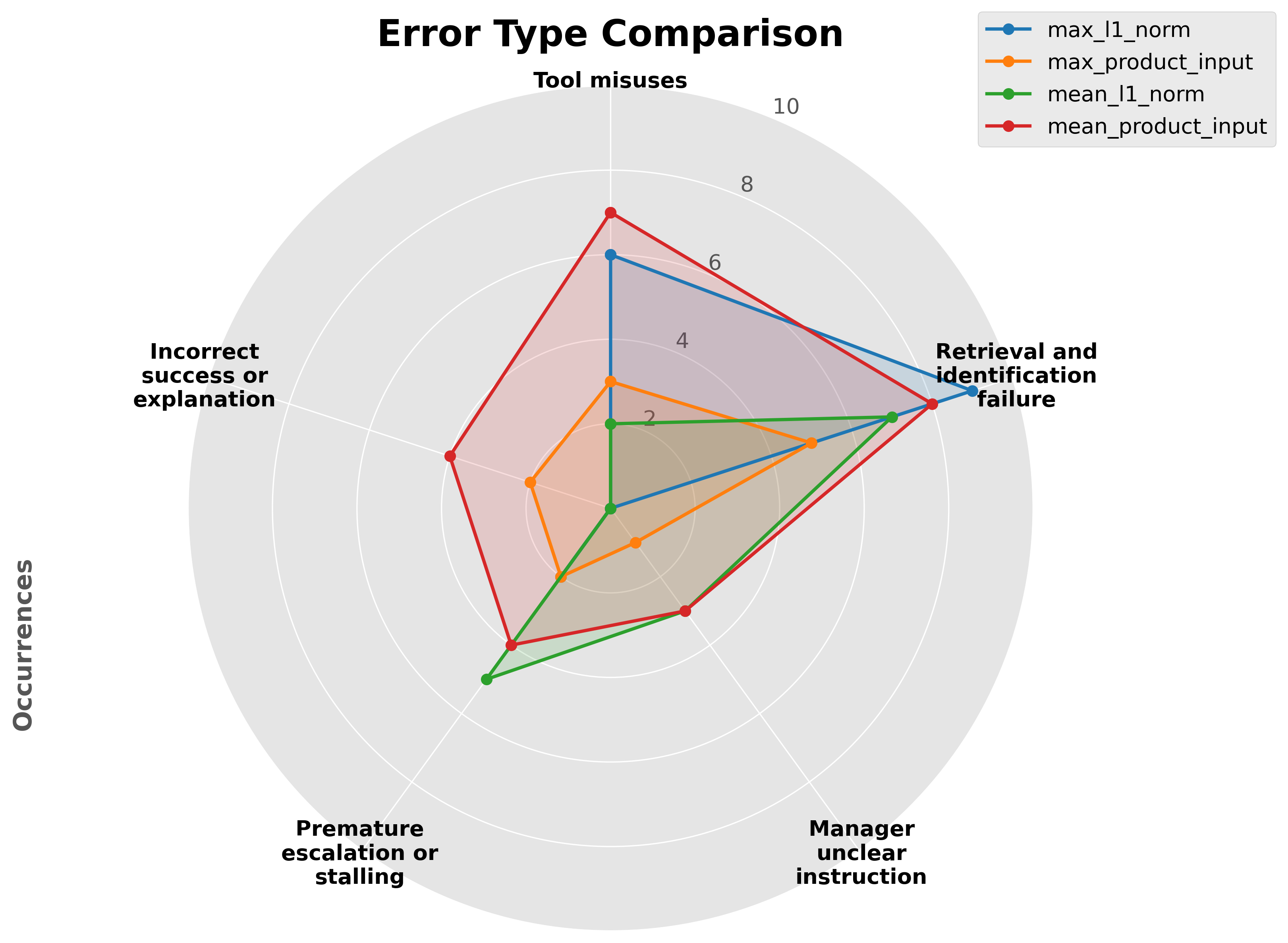}
    \caption{The occurrences of different error types detected by the optimizer under different connection weight formulae on $\tau$-bench with Qwen-3-32B.}
    \label{fig:taubench error analysis}
\end{figure}

\paragraph{Error Analysis}
We identify five error types: tool misuse, retrieval/identification failure, unclear manager instructions, premature escalation, and incorrect explanations.

As shown in Figure~\ref{fig:taubench error analysis}, retrieval and identification failures dominate across connection-weight settings. This reflects a central difficulty of $\tau$-bench: successful task completion depends not only on selecting the right tool, but also on resolving the correct user, order, and task state across multiple turns. When this grounding step fails, later tool calls and final responses are likely to become incorrect even if the overall workflow is reasonable. Tool misuse also contributes to failures, indicating that action selection remains important. Premature escalation or stalling shows that agents sometimes fail to persist through the required tool-use procedure, while incorrect success or explanation affects whether the final response communicates the correct outcome. Overall, these errors highlight the long-horizon nature of $\tau$-bench, where reliable retrieval, entity tracking, and inter-agent information transfer are necessary for both high action reward and high output reward.

%% file: tables/multiwoz_main.tex
\begin{table}[h]
    \centering

        \scriptsize
        \setlength{\tabcolsep}{3pt}
        
        \begin{tabular}{l | c c c c c c} \toprule
            \textbf{Model} & \textbf{Inform} & \textbf{Success} & \textbf{JGA} & \textbf{Recall} & \textbf{Precision} & \textbf{Slot F1} \\
            \midrule[1pt]
            
            \multicolumn{7}{c}{\textbf{Single-Agent}} \\ 
            \midrule[1pt]
            
            Llama-3.3-70B-It & 84.0 & 71.0 & 40.3 & 96.9 & 81.5 & 88.5 \\
            Qwen-3-32B & 88.0 & 40.0 & 44.4 & 97.7 & 80.9 & 88.5 \\
            \midrule[1pt]
            
            \multicolumn{7}{c}{\textbf{Multi-Agent (Before Optimization)}} \\ 
            \midrule[1pt]
            
            Llama-3.3-70B-It & 87.0 & 57.0 & 24.3 & 97.4 & 54.1 & 69.5 \\
            Qwen-3-32B & 95.0 & 80.0 & 28.9 & 92.5 & 69.4 & 79.3 \\
            \midrule[1pt]

            \multicolumn{7}{c}{\textbf{Multi-Agent (Optimized with Mean of L1 Norm)}} \\ 
            \midrule[1pt]
            
            Llama-3.3-70B-It & 42.0 & 07.0 & 28.6 & 90.7 & 78.0 & 83.9 \\
            Qwen-3-32B & \textbf{\underline{99.0}} & \textbf{94.0} & \textbf{\underline{54.4}} & \textbf{98.1} & \textbf{\underline{85.5}} & \textbf{\underline{91.4}} \\
            \midrule[1pt]

            \multicolumn{7}{c}{\textbf{Multi-Agent (Optimized with Max of L1 Norm)}} \\ 
            \midrule[1pt]
            
            Llama-3.3-70B-It & 87.0 & 54.0 & 36.7 & 94.5 & 78.9 & 86.0 \\
            Qwen-3-32B & \textbf{\underline{99.0}} & \textbf{\underline{95.0}} & \textbf{53.0} & \textbf{\underline{98.3}} & \textbf{85.2} & \textbf{91.3} \\
            \midrule[1pt]

            \multicolumn{7}{c}{\textbf{Multi-Agent (Optimized with Mean of Product with Input)}} \\ 
            \midrule[1pt]
            
            Llama-3.3-70B-It & 82.0 & 59.0 & 39.0 & 96.7 & 79.6 & 87.4 \\
            Qwen-3-32B & 98.0 & 83.0 & 51.9 & 96.9 & 84.6 & 90.3 \\
            \midrule[1pt]
            
            \multicolumn{7}{c}{\textbf{Multi-Agent (Optimized with Max of Product with Input)}} \\ 
            \midrule[1pt]
            
            Llama-3.3-70B-It & 85.0 & 62.0 & 38.9 & 94.7 & 81.7 & 87.7 \\
            Qwen-3-32B & 95.0 & 86.0 & 51.0 & 97.4 & 79.3 & 87.4 \\
            \bottomrule
        \end{tabular}

    \hfill
    \caption{Single-agent System and multi-agent system performance on MultiWOZ 2.4. The table shows the inform score, success score, joint goal accuracy (JGA), slot recall, slot precision, and slot F1 score. \textbf{\underline{Best results}} are bold and underlined; \textbf{second-best} are bold.}
    \label{tab:multiwoz_main}
\end{table}

%% file: tables/taubench_main.tex
\begin{table}[h]
    \centering

        \scriptsize
        \setlength{\tabcolsep}{3.5pt}
        
        \begin{tabular}{l | c c c} \toprule
            \textbf{Model} & \textbf{Action Reward} & \textbf{Output Reward} & \textbf{Overall Reward} \\
            \midrule[1pt]
            
            \multicolumn{4}{c}{\textbf{Single-Agent}} \\ 
            \midrule[1pt]
            
            Llama-3.3-70B-It & 09.6 & 62.6 & 07.0 \\
            Qwen-3-32B & \textbf{27.8} & 71.3 & 22.6 \\
            \midrule[1pt]
            
            \multicolumn{4}{c}{\textbf{Multi-Agent (Before Optimization)}} \\ 
            \midrule[1pt]
            
            Llama-3.3-70B-It & 09.6 & 70.4 & 06.1 \\
            Qwen-3-32B & 13.9 & \textbf{78.3} & 13.0 \\
            \midrule[1pt]

            \multicolumn{4}{c}{\textbf{Multi-Agent (Optimized with Mean of L1 Norm)}} \\ 
            \midrule[1pt]
            
            Llama-3.3-70B-It & 11.3 & 70.4 & 08.7 \\
            Qwen-3-32B & 17.4 & 73.0 & 13.9 \\
            \midrule[1pt]

            \multicolumn{4}{c}{\textbf{Multi-Agent (Optimized with Max of L1 Norm)}} \\ 
            \midrule[1pt]
            
            Llama-3.3-70B-It & 13.9 & 73.0 & 09.6 \\
            Qwen-3-32B & \textbf{\underline{28.7}} & \textbf{\underline{79.1}} & \textbf{\underline{24.3}} \\
            \midrule[1pt]

            \multicolumn{4}{c}{\textbf{Multi-Agent (Optimized with Mean of Product with Input)}} \\ 
            \midrule[1pt]
            
            Llama-3.3-70B-It & 12.2 & 72.2 & 08.7 \\
            Qwen-3-32B & 27.0 & \textbf{78.3} & \textbf{\underline{24.3}} \\
            \midrule[1pt]
            
            \multicolumn{4}{c}{\textbf{Multi-Agent (Optimized with Max of Product with Input)}} \\ 
            \midrule[1pt]
            
            Llama-3.3-70B-It & 13.9 & 71.3 & 09.6 \\
            Qwen-3-32B & 20.0 & 77.4 & 17.4 \\
            \bottomrule
        \end{tabular}

    \hfill
    \caption{Single-agent system and multi-agent system performance on $\tau$-bench. The table shows the action rewards, output rewards, and overall rewards for the retail domain. \textbf{\underline{Best results}} are bold and underlined; \textbf{second-best} are bold.}
    \label{tab:taubench_main}
\end{table}

%% file: sections/conclusion.tex
\section{Conclusion}

In this work, we introduce Gradient-Based Connections (GBC), a novel framework for fine-grained attribution and optimization in multi-agent systems. By modeling a multi-agent workflow as a computational graph and leveraging gradient-based signals at the token level, GBC enables precise identification of how intermediate agent outputs influence downstream decisions. This formulation addresses a fundamental limitation of prior approaches—namely, the lack of effective credit assignment across agents—and provides a principled mechanism for diagnosing and improving multi-agent coordination.

Building on GBC, we develop AgentChord, an efficient optimization framework that integrates attribution-guided feedback with iterative prompt refinement. Through a prefix-based gradient computation strategy, AgentChord makes gradient-based attribution feasible for large-scale language models. Empirical results on MultiWOZ and $\tau$-bench demonstrate that GBC consistently improves multi-agent performance across a range of metrics, and in many cases enables multi-agent systems to match or surpass strong single-agent baselines. Analysis of the MultiWOZ results further reveals that higher attribution quality is associated with greater optimization effectiveness. Furthermore, our analysis provides meaningful insights into error patterns, revealing both system-level weaknesses and task-intrinsic challenges.

Overall, our work highlights the importance of token-level, cross-agent credit assignment as a key component for advancing multi-agent systems. We believe GBC offers a general and extensible approach for future research on interpretable and optimizable multi-agent architectures.

%% file: sections/limitations.tex
\section*{Limitations}

Despite its effectiveness, our approach has several limitations that warrant further investigation.

First, computational cost remains a concern. Although the prefix-based optimization reduces memory overhead, gradient-based attribution still requires multiple forward and backward passes through LLMs, making the approach expensive compared to purely black-box methods. This constraint limits scalability to larger systems or longer interaction horizons.

Second, GBC relies on the quality and design of the verbal loss function. Since the loss is task-specific and expressed in natural language, its effectiveness depends on how well it captures fine-grained errors. Poorly designed loss signals may lead to noisy or misleading attribution, reducing optimization effectiveness.

Third, while GBC provides token-level attribution, it still operates under a first-order approximation of influence (e.g., gradient-based signals). This may not fully capture complex nonlinear interactions between agents, especially in long multi-turn or highly entangled workflows.

Fourth, our experiments focus on specific benchmarks (MultiWOZ and $\tau$-bench) and particular system architectures (e.g., manager–worker setups). Although results are promising, the generalization of GBC to other domains—such as open-ended reasoning, code generation, or large-scale autonomous agent systems—remains to be validated.

Finally, our analysis reveals that some error types—such as cross-domain errors, information omission, and retrieval failures—persist even after optimization, suggesting that these may stem from intrinsic task difficulty or limitations of current LLMs, rather than attribution alone.

Future work may explore more efficient gradient approximations, improved verbal loss design, integration with reinforcement learning, and extensions to dynamic or adaptive multi-agent topologies.

%% file: sections/appendix.tex
\section{Experiment Setting Detail}
Unless otherwise specified, all experiments are conducted on a single compute node equipped with four NVIDIA A40 GPUs, 208 GB of system memory, and 16 CPU cores. We use the same hardware configuration for both the optimization and inference phases to ensure consistency across experimental runs. For local model serving, we enable FP8 quantization whenever supported by the target model and serving backend. We use GPT-4 as the optimizer model for prompt refinement throughout the optimization process. When using Qwen-3-32B, we turn off the thinking process in order to accelerate the experiments.


\section{Verbal Loss Prompts}
\label{app:loss}
This section presents the verbal-loss prompt templates used for different task scenarios, including MultiWOZ and $\tau$-bench.

\subsection{Verbal Loss Prompts for MultiWOZ}
\label{app:multiwoz loss}
For MultiWOZ, we design two types of verbal loss: Joint Goal Accuracy (JGA) loss and Inform \& Success loss. JGA loss is computed at the turn level. We present the prompt templates for both losses below.

\subsubsection{JGA Loss}
\label{app:jga loss}
The prompt template for JGA loss is shown below:

\begin{tcolorbox}[
    colback=gray!3!white,
    colframe=black!30!white, 
    title={Prompt Template for JGA Loss},
    fonttitle=\bfseries,
    boxrule=0.5pt,
    arc=4pt,
    boxsep=5pt,
    left=6pt,
    right=6pt,
    top=6pt,
    bottom=6pt,
    coltitle=black,
    breakable
]
The system has made the following user intention predictions:
\begin{lstlisting}[basicstyle=\ttfamily\small, breaklines=true]
{prediction}
\end{lstlisting}
The ground truth user intentions are:
\begin{lstlisting}[basicstyle=\ttfamily\small, breaklines=true]
{ground_truth}
\end{lstlisting}
The false positive predictions are:
\begin{lstlisting}[basicstyle=\ttfamily\small, breaklines=true]
{false_positive}
\end{lstlisting}
The false negative predictions are:
\begin{lstlisting}[basicstyle=\ttfamily\small, breaklines=true]
{false_negative}
\end{lstlisting}
\end{tcolorbox}

When using this template, "prediction" is replaced with the JSON string of the predicted dialogue state for a given turn. Similarly, "ground\_truth", "false\_positive", and "false\_negative" are replaced with the JSON strings of the ground-truth dialogue state, false-positive slot-value pairs, and false-negative slot names, respectively. An example dialogue state is shown below:
\begin{lstlisting}[basicstyle=\ttfamily\small, breaklines=true]
{
    "train-departure": "cambridge", 
    "train-leaveAt": "11:00", 
    "train-day": "wednesday", 
    "train-destination": "stansted airport"
}
\end{lstlisting}

\subsubsection{Inform \& Success Loss}
\label{app:inf&suc loss}
The prompt template for Inform \& Success loss is shown below:

\begin{tcolorbox}[
    colback=gray!3!white,
    colframe=black!30!white, 
    title={Prompt Template for Inform \& Success Loss},
    fonttitle=\bfseries,
    boxrule=0.5pt,
    arc=4pt,
    boxsep=5pt,
    left=6pt,
    right=6pt,
    top=6pt,
    bottom=6pt,
    coltitle=black,
    breakable
]
The system has made the following queries:
\begin{lstlisting}[basicstyle=\ttfamily\small, breaklines=true]
{provided_queries}
\end{lstlisting}
The ground truth queries are:
\begin{lstlisting}[basicstyle=\ttfamily\small, breaklines=true]
{requested_queries}
\end{lstlisting}
The system has provided the following information:
\begin{lstlisting}[basicstyle=\ttfamily\small, breaklines=true]
{provided_information}
\end{lstlisting}
The ground truth information is:
\begin{lstlisting}[basicstyle=\ttfamily\small, breaklines=true]
{requested_information}
\end{lstlisting}
\end{tcolorbox}

When using this template, "provided\_queries" and "requested\_queries" are replaced with the JSON strings of the queries generated by the system and the corresponding ground-truth queries. An example is shown below:
\begin{lstlisting}[basicstyle=\ttfamily\small, breaklines=true]
{
    "hotel": [
        {
            "area": "east", 
            "internet": "yes", 
            "parking": "no"
        }, 
        {
            "area": "east", 
            "internet": "yes", 
            "parking": "no"
        }, 
        {
            "area": "east", 
            "internet": "yes", 
            "parking": "no"
        }
    ]
}
\end{lstlisting}

Similarly, "provided\_information" and "requested\_information" are replaced with the information provided by the system and the corresponding ground-truth information. An example is shown below:
\begin{lstlisting}[basicstyle=\ttfamily\small, breaklines=true]
{
    "attraction": ["POST"], 
    "taxi": ["PHONE"]
}
\end{lstlisting}

\subsection{Verbal Loss Prompt for $\tau$-bench}
\label{app:reward loss}
The prompt template for reward loss is shown below:

\begin{tcolorbox}[
    colback=gray!3!white,
    colframe=black!30!white, 
        title={Prompt Template for Reward Loss},
    fonttitle=\bfseries,
    boxrule=0.5pt,
    arc=4pt,
    boxsep=5pt,
    left=6pt,
    right=6pt,
    top=6pt,
    bottom=6pt,
    coltitle=black,
    breakable
]
The ground truth actions are:
\begin{lstlisting}[basicstyle=\ttfamily\small, breaklines=true]
{groundtruth_actions}
\end{lstlisting}
The predicted actions are:
\begin{lstlisting}[basicstyle=\ttfamily\small, breaklines=true]
{predicted_actions}
\end{lstlisting}
Whether the actions match:
\begin{lstlisting}[basicstyle=\ttfamily\small, breaklines=true]
{action_match}
\end{lstlisting}
The required text strings in the responses are:
\begin{lstlisting}[basicstyle=\ttfamily\small, breaklines=true]
{required_output_strings}
\end{lstlisting}
The predicted responses are:
\begin{lstlisting}[basicstyle=\ttfamily\small, breaklines=true]
{predicted_responses}
\end{lstlisting}
Whether the responses match:
\begin{lstlisting}[basicstyle=\ttfamily\small, breaklines=true]
{output_match}
\end{lstlisting}
\end{tcolorbox}

When using this template, "groundtruth\_actions" and "predicted\_actions" are replaced with the ground-truth tool-call trajectories and the generated trajectories, respectively. An example is shown below:
\begin{lstlisting}[basicstyle=\ttfamily\small, breaklines=true]
[
    {
        "tool_name": "exchange_delivered_order _items",
        "tool_arguments": {
            "order_id": "#W9077205",
            "item_ids": ["3877338112"],
            "new_item_ids": ["2444431651"],
            "payment_method_id": "gift_card_7108145"
        }
    }
]
\end{lstlisting}

Similarly, "action\_match" and "output\_match" are replaced with the corresponding binary rewards, while "required\_output\_strings" and "predicted\_responses" are replaced with the list of required text strings and the list of system utterances, respectively.

\section{Language Model as Optimizer}
\label{app:opro}
\subsection{Pseudocode of Language Model as Optimizer}
Algorithm~\ref{alg:lm_optimizer} presents the pseudocode for using a language model as an optimizer.
\input{algorithms/optimizer}

\subsection{Prompt of Language Model as Optimizer}
At each optimization step, the optimizer receives an instruction prompt and an input prompt. We present the corresponding prompt templates below. The instruction prompt is as follows:

\begin{tcolorbox}[
    colback=gray!3!white,
    colframe=black!30!white, 
    title={Prompt for Optimizer Instruction},
    fonttitle=\bfseries,
    boxrule=0.5pt,
    arc=4pt,
    boxsep=5pt,
    left=6pt,
    right=6pt,
    top=6pt,
    bottom=6pt,
    coltitle=black,
    breakable
]

You are an expert in analyzing multi-agent systems and providing insights on how to improve agent performance through better prompts.\\
You will be given the structure of a multi-agent system, including the names of different agents and their current prompts.\\
You will be provided with some inference trajectories from the multi-agent system. Each trajectory consists of a sequence of outputs from different agents. Each output is conditioned on the previous outputs.\\
At the end of each trajectory, there is a comparison between the final output and the expected output.\\
You will also receive an optimization history that contains information about the agents and their prompts.\\
Your task is to analyze these trajectories and provide insights on how the agents can avoid the mistakes and achieve better results by improving their prompts. Note that the order of the agents is fixed. So you should not suggest changing the order of the agents.\\
If there's no **Warning** section in the prompt, you can add such a section at the end of the prompt. You can add sentences to warn the agents about the mistakes in the trajectories. For example, if an agent often misses certain tool calls, you can encourage the agent to use certain tools under certain situations; if an agent often misses certain information in tool call inputs, you can warn the agent to pay attention to that information.\\
You should format the warnings in bullet points in markdown format. For each warning, you should attach a failure case from the trajectories as an example.\\
Note that the toolkit of each agent might be different. So you should not suggest using tools that are not available to the agent.\\
However, the content within the \{DONT\_CHANGE\_HEADER\} and \{DONT\_CHANGE\_FOOTER\} tags and the tags themselves should be kept unchanged and preserved in the new prompt.\\
You should also note that the causality in some trajectories is noisy. So you should not take the noisy causality into account.\\

\# Procedure\\
You should follow these steps:
1. Identify the agents that need to improve their prompts.
2. For each selected agent, suggest a new prompt that could lead to better results.

\# Input Format\\
You will receive the optimization history and trajectories in the following format:
The structure of the multi-agent system:
\begin{lstlisting}[basicstyle=\ttfamily\small, breaklines=true]
{
  "agent_name_1": "Current prompt for agent 1",
  "agent_name_2": "Current prompt for agent 2",
  ...
}
\end{lstlisting}
The tools available to each agent:
\begin{lstlisting}[basicstyle=\ttfamily\small, breaklines=true]
{
  "agent_name_1": [... tool descriptions ...],
  "agent_name_2": [... tool descriptions ...],
  ...
}
\end{lstlisting}
The optimization history and the corresponding performance:
\begin{lstlisting}[basicstyle=\ttfamily\small, breaklines=true]
[
  {
    "prompts": {
      "agent_name_1": "Prompt for agent 1",
      "agent_name_2": "Prompt for agent 2",
      ...
    },
    "performance": "A brief description of the performance of the multi-agent system, including the expected output and the actual output.",
  },
  {
    "prompts": {
      "agent_name_1": "Prompt for agent 1",
      "agent_name_2": "Prompt for agent 2",
      ...
    },
    "performance": "A brief description of the performance of the multi-agent system, including the expected output and the actual output.",
  },
  ...
]
\end{lstlisting}
The inference trajectories:
\begin{lstlisting}[basicstyle=\ttfamily\small, breaklines=true]
[
  "agent_1: output_1 -> agent_2: output_2 -> ... -> agent_n: final_output -> loss: comparison_of_final_output_and _expected_output",
  "agent_1: output_1 -> agent_2: output_2 -> ... -> agent_n: final_output -> loss: comparison_of_final_output_and _expected_output",
  ...
]
\end{lstlisting}

\# Output Format
You should output a JSON object with the following structure:
\begin{lstlisting}[basicstyle=\ttfamily\small, breaklines=true]
{
  "reasoning": "Your reasoning about the agents and their prompts.",
  "agent_prompts": {
    "agent_name_x": "New prompt for agent x containing the **Warning** section.",
    ...
  }
}
\end{lstlisting}
\end{tcolorbox}

The input prompt template is as follows:

\begin{tcolorbox}[
    colback=gray!3!white,
    colframe=black!30!white, 
    title={Prompt Template for Inform \& Success Loss},
    fonttitle=\bfseries,
    boxrule=0.5pt,
    arc=4pt,
    boxsep=5pt,
    left=6pt,
    right=6pt,
    top=6pt,
    bottom=6pt,
    coltitle=black,
    breakable
]
The structure of the multi-agent system:
\begin{lstlisting}[basicstyle=\ttfamily\small, breaklines=true]
{agent_structure}
\end{lstlisting}
The tools available to each agent:
\begin{lstlisting}[basicstyle=\ttfamily\small, breaklines=true]
{agent_tools}
\end{lstlisting}
The optimization history and the corresponding performance:
\begin{lstlisting}[basicstyle=\ttfamily\small, breaklines=true]
{optimization_history}
\end{lstlisting}
The inference trajectories:
\begin{lstlisting}[basicstyle=\ttfamily\small, breaklines=true]
{inference_trajectories}
\end{lstlisting}
\end{tcolorbox}

\section{Time Cost}

Table~\ref{tab:runtime_connection_weight} reports the running time of completing 10 steps of optimization under different connection weight settings on MultiWOZ and $\tau$-bench. Overall, Llama-3.3-70B-It requires substantially longer running time than Qwen-3-32B across both benchmarks, which is expected given the larger model size. On MultiWOZ, the running times are relatively stable across different connection weight choices: Llama-3.3-70B-It takes around 16.3--16.7 hours, while Qwen-3-32B takes around 8.0--9.5 hours. On $\tau$-bench, the runtime varies more noticeably for Llama-3.3-70B-It, ranging from 10.3 to 16.3 hours depending on the connection weight, whereas Qwen-3-32B remains consistently around 5--6 hours. These results indicate that the choice of connection weight does not introduce a substantial additional computational burden, and the overall runtime is primarily determined by the benchmark and the underlying model.

\input{tables/time_cost}

%% file: algorithms/optimizer.tex
\begin{algorithm}[h]
\small
\caption{Language Model as Optimizer}
\label{alg:lm_optimizer}
\begin{algorithmic}[1]
\Require Initial agent prompts $P^{0}$
\Require agent tools $\mathcal{A}$
\Require Prompt template $\mathcal{F}(\cdot)$
\State Initialize agent prompts $P \leftarrow P^{(0)}$
\State Initialize optimization history $H \leftarrow [\ ]$
\For{$t = 1$ to $T$}
    \State Run the MAS on the training samples and collect latest prompts $P^{t}$, latest performance $r^{t}$, and latest attribution trajectories $\mathcal{T}^{t}$
    \State $P \leftarrow P^{t}$
    \State $r \leftarrow r^{t}$
    \State $\mathcal{T} \leftarrow \mathcal{T}^{t}$
    \State Append $(P, r)$ to optimization history $H$
    \State $M \leftarrow \mathcal{F}(P, \mathcal{A}, H, \mathcal{T})$
    \State Query LLM with $M$ to obtain prompt updates $\Delta P$
    \State Update prompts: $P \leftarrow P + \Delta P$
\EndFor
\State \Return $P$
\end{algorithmic}
\end{algorithm}

%% file: tables/time_cost.tex
\begin{table}[h]
    \centering

    \scriptsize
    \setlength{\tabcolsep}{9pt}

    \begin{tabular}{l | l c}
        \toprule
        \textbf{Model} & \textbf{Connection Weight} & \textbf{Time} \\
        \midrule[1pt]

        \multicolumn{3}{c}{\textbf{MultiWOZ}} \\
        \midrule[1pt]

        \multirow{4}{*}{Llama-3.3-70B-It}
            & Mean of L1 Norm & \texttt{16h39m20s} \\
            & Max of L1 Norm & \texttt{16h40m46s} \\
            & Mean of Product with Input & \texttt{16h26m19s} \\
            & Max of Product with Input & \texttt{16h19m55s} \\

        \midrule

        \multirow{4}{*}{Qwen-3-32B}
            & Mean of L1 Norm & \texttt{08h45m25s} \\
            & Max of L1 Norm & \texttt{08h00m16s} \\
            & Mean of Product with Input & \texttt{09h27m18s} \\
            & Max of Product with Input & \texttt{08h26m49s} \\

        \midrule[1pt]

        \multicolumn{3}{c}{\textbf{$\tau$-bench}} \\
        \midrule[1pt]

        \multirow{4}{*}{Llama-3.3-70B-It}
            & Mean of L1 Norm & \texttt{16h20m07s} \\
            & Max of L1 Norm & \texttt{12h52m50s} \\
            & Mean of Product with Input & \texttt{14h47m31s} \\
            & Max of Product with Input & \texttt{10h16m37s} \\

        \midrule

        \multirow{4}{*}{Qwen-3-32B}
            & Mean of L1 Norm & \texttt{05h09m28s} \\
            & Max of L1 Norm & \texttt{05h59m47s} \\
            & Mean of Product with Input & \texttt{05h06m50s} \\
            & Max of Product with Input & \texttt{06h04m16s} \\

        \bottomrule
    \end{tabular}

    \caption{Running time of optimization under different connection weight settings on MultiWOZ and $\tau$-bench.}
    \label{tab:runtime_connection_weight}
\end{table}